# Performance Analysis of the Globus Toolkit Monitoring and Discovery Service, MDS2

Xuehai Zhang
*Department of Computer Science*
*University of Chicago*
hai@cs.uchicago.edu

Jennifer M. Schopf
*Mathematics and Computer Science Division*
*Argonne National Laboratory*
jms@mcs.anl.gov

## Abstract

*Monitoring and information services form a key component of a distributed system, or Grid. A quantitative study of such services can aid in understanding the performance limitations, advise in the deployment of the monitoring system, and help evaluate future development work. To this end, we examined the performance of the Globus Toolkit® Monitoring and Discovery Service (MDS2) by instrumenting its main services using NetLogger. Our study shows a strong advantage to caching or prefetching the data, as well as the need to have primary components at well-connected sites.*

## 1. Introduction

Grid platforms[3] depend on monitoring and information services to support the discovery and monitoring of the distributed resources for various tasks. Indepth studies are needed to understand any performance limitations in common settings.

In our previous work[13], we investigated the behaviors of the Globus Toolkit Monitoring and Discovery Service (MDS2)[2][9], the most common monitoring system currently used for production Grids, with the focus on analyzing the end-to-end performance of a user request at a very coarse grain. To better understand the unexplained behaviors we saw in that study, in this work we exam MDS behavior at a finer granularity by using NetLogger[11][12] technologies to instrument both MDS2 server and client codes and running experiments to evaluate the effect of a large number of concurrent users accessing the different services.

## 2. MDS2

The Monitoring and Discovery Service (MDS2)[2][9] is the Grid information service used in the Globus Toolkit[6]. MDS2, built on top of the Lightweight Directory Access Protocol (LDAP) (v3)[10], is used primarily to address the resource selection problem, namely, how a user identifies the host or set of hosts on which to run an application. MDS2 provides a uniform, flexible interface to data collected by lower-level information providers. It has a decentralized structure that allows it to scale, and it can handle static or dynamic data.

**Table 1: Definitions of the seven phases of an MDS2 query.**

| Phase Name | Phase Definition | Instrumentation Location |
|---|---|---|
| *Client-Connect* | Stage for MDS client program to open a connection to MDS server | Client side |
| *Client-Bind* | Stage for MDS client to authenticate to MDS server | Client side |
| *Server-InitSearch* | Stage for MDS server performs search initialization | Server side |
| *Server-SearchIndex* | Stage for MDS server search indexes for entries | Server side |
| *Server-Invoking* | Stage for MDS server to invoke reported information providers or GRIS to generate fresh data | Server side |
| *Server-GenResult* | Stage for MDS server to build the results | Server side |
| *Client-EndConnect* | Stage for MDS client to receive results and disconnect | Client side |

MDS2 has a hierarchical structure that consists of three main components. A Grid Index Information Service (GIIS) provides an aggregate directory of lower-level data. A Grid Resource Information Service (GRIS) runs on a resource and acts as a modular content gateway for a resource. Information providers (IPs) interface from any data collection service and then talk to a GRIS. Each service registers with high-level services using a soft-state protocol that allows dynamic cleaning of dead resources. Each level also has caching to minimize the transfer of unstale data and lessen network overhead.

We use NetLogger to instrument both the MDS2 server and client codes. NetLogger[11][12] is a toolkit developed by Lawrence Berkley National Laboratory to monitor, under actual operating conditions, the behavior of elements of a complex distributed system in order to determine exactly where time is spent within such a system and identify the performance bottlenecks. With NetLogger, the components of a distributed system can be modified to produce time-stamped logs of "interesting" events at all the critical points of the system, which are then correlated to allow the characterization of the performance of all aspects of the system in detail. To instrument an application to produce event logs, the application developer inserts calls to the NetLogger API at all the critical points in the code, then links the application with the NetLogger library. NetLogger is a lightweight tool and adds little overhead to existing program when used carefully[11].

By adding NetLogger calls we broke the end-to-end path of a MDS2 request into seven phases: (1) *Client-Connect*, (2) *Client-Bind*, (3) *Server-InitSearch*, (4) *Server-SearchIndex*, (5) *Server-Invoking*, (6) *Server-GenResult*, and (7) *Client-EndConnect*, as shown in Table 1. Phases 1, 2, and 7 constitute the MDS2 client side components, and phases 3–6 constitute the server-side components. A NetLogger view of the behavior of a MDS v2.4 GRIS without data caching accessed by 10 concurrent users is given in Figure 1.

## 3. MDS2 Performance Results

In this section, we discuss the experiments results and evaluations for MDS2. First we briefly talk about experimental setup, and then we describe the metrics we used in the experiments. Finally we present the experiments results and analysis.

### 3.1. Experimental Setup

We ran our experiments between two sites: the Lucky testbed at Argonne National Laboratory (ANL), which provided the MDS2 server-side services, and a testbed at the University of Chicago (UC), which provided the client-side services.

The Lucky testbed includes seven Linux machines with hostnames *lucky{0,1,3,..,7}.mcs.anl* (*lucky2* was unavailable during the experiments) and a shared file system on a 100 Mbps LAN. Each machine is equipped with two 1133 MHz Intel PIII CPUs (with a 512 KB cache per CPU) and 512 MB RAM. *Lucky0* and *lucky6* run Linux kernel 2.4.10 and the rest run kernel 2.4.19.

The UC client-side hosts are a cluster of 20 Linux machines with a shared file system on a 100 Mbps LAN. Fifteen of them were equipped with a 1208 MHz CPU and 256 MB RAM, while the rest had a slightly slower CPU (but at least 756 MHz), also with 256 MB RAM. Each machine runs Linux kernel 2.4.17 or a higher version.

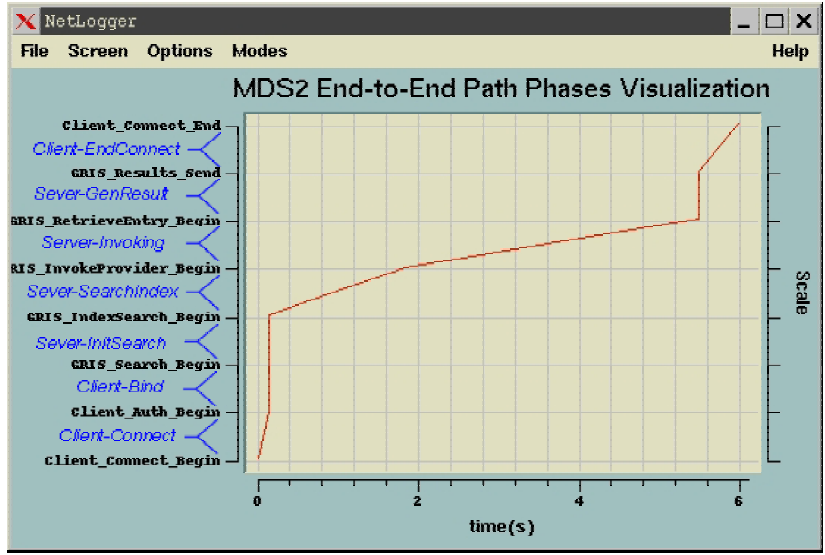

**Figure 1: The seven phases of an MDS2 query, as shown using NetLogger instrumentation.**

The bandwidth between ANL and UC is around 55 Mbits per sec on average (as measured by Iperf[8]), and the latency (Round-Trip Time) is approximately 2.3 msec on average.

We deployed MDS 2.2 and 2.4 on both sites and used NetLogger v2.0.13 to instrument the server and client codes of both versions. To synchronize the clock, we ran NTP 4.1.2 at both the Lucky testbed and UC client hosts.

In our experiments, we simulated up to 600 users querying the MDS2 services simultaneously for 10 minutes, with a waiting period of one second between receiving a request response and issuing the next response, by running individual user processes (scripts) on client machines. We selected 600 as the upper bound since the average number of concurrent users for MDS2 services in real world is much smaller. We evenly distributed the simulated users to all twenty machines to balance the load.

We used Ganglia[5], a cluster monitoring system developed by UC Berkeley, to collect the performance data at five-second intervals. The values reported in each experiment are the average over all the values recorded during a 10-minute time span. We performed all the experiments in a LAN setting to ensure that the performance of the service was affected primarily by the service components and not by other external factors.

### 3.2. Performance Metrics

The performance metrics we used in our work include throughput, observed response time (*ORT*), request processing time (*RPT*), load1, and CPU-load.

Throughput is defined as the average number of requests (or queries) processed by a MDS2 service component per second.

*ORT*, equivalent to the metric response time used in our previous work[13], denotes the average amount of time (in seconds) from the point a user sends out a request till the user gets the response back. It is calculated at the client side. *RPT* is defined as the average time spent at the server side for a MDS2 service to handle a user request. *ORT* is always greater than *RPT*, and their relationship can be represented by

$$ORT = T_{Client\text{-}Connect} + T_{Client\text{-}Bind} + RPT + T_{Client\text{-}EndConnect} \quad (1)$$

where $T_{Client\text{-}Connect}$, $T_{Client\text{-}Bind}$ and $T_{Client\text{-}EndConnect}$ denote the time spent on the *Client-Connect phrase,* the *Client-Bind* phrase and the *Client-EndConnect* phase respectively. As shown in Table 1, the server side consists of four phases that timewise result in *RPT*. Therefore, Equation 1 can be expanded to

$$ORT = T_{Client\text{-}Connect} + T_{Client\text{-}Bind} + RPT + T_{Client\text{-}EndConnect}$$
$$RPT = T_{Server\text{-}InitSearch} + T_{Server\text{-}SearchIndex} + T_{Server\text{-}Invoking} + T_{Server\text{-}GenResult} \quad (2)$$

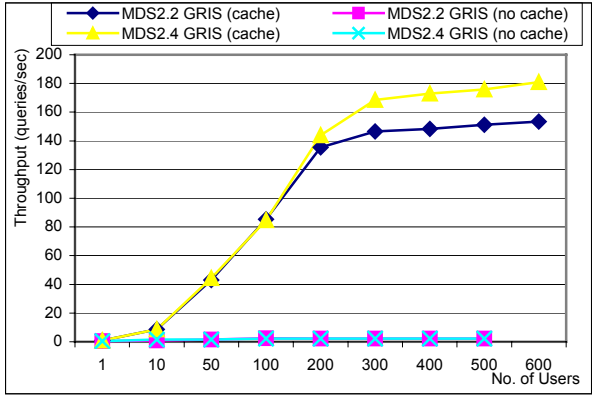

**Figure 2: MDS 2.2 and 2.4 GRIS Throughput vs. No. of Concurrent Users**

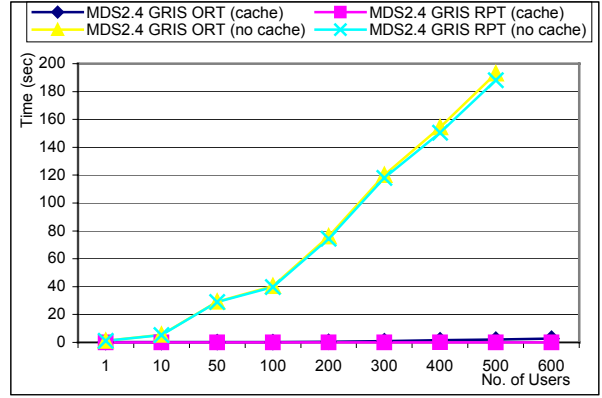

**Figure 3: MDS 2.4 GRIS *ORT* and *RPT* vs. No. of Concurrent Users**

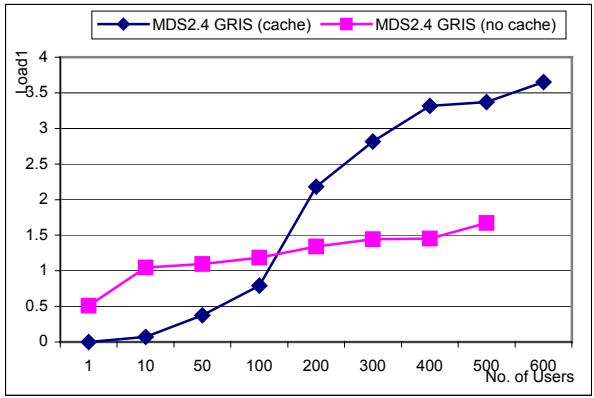

**Figure 4: MDS 2.4 GRIS Host Load1 vs. No. of Concurrent Users**

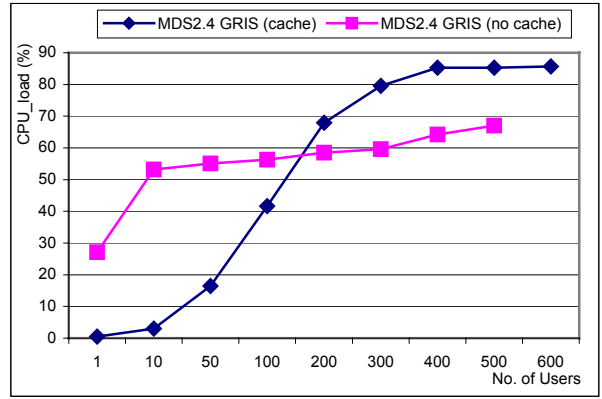

**Figure 5: MDS 2.4 GRIS Host CPU Load vs. No. of Concurrent Users**

We also used two load metrics for the experiments, a one-minute load average (load1) and CPU-load. The metricLoad1 is the average number of processes in the ready queue waiting to run over the last minute measured by the Ganglia metric "load_one". Usually the system is overloaded if the load1 value is greater than 3. CPU-load indicates the percentage of the CPU cycles spent in user mode and system mode, which we measured by averaging the sum of cpu_user and cpu_system recorded by Ganglia.

may be low while load1 is high if the same machine is trying to run a large number of applications that are blocking on I/O.

### 3.3. MDS2 Information Server Scalability

As the information server of MDS2, the GRIS can be heavily queried by users. Therefore, in our first set of experiments we evaluated its performance when it is accessed by a large number of users concurrently.

For each MDS2 version, we ran a GRIS on *lucky7*, which had ten information providers reporting to it. We examined two different scenarios: the GRIS always caching the data from the information providers and the GRIS never caching the data. Our intention was to understand the GRIS performances under two extreme conditions, in order to help us estimate the performance of the average case, which is somewhere between these two. Each query requested all the data elements in the GRIS directory, and this data was generated by all the reported information providers. The average size of requested data was less than 10 KB.

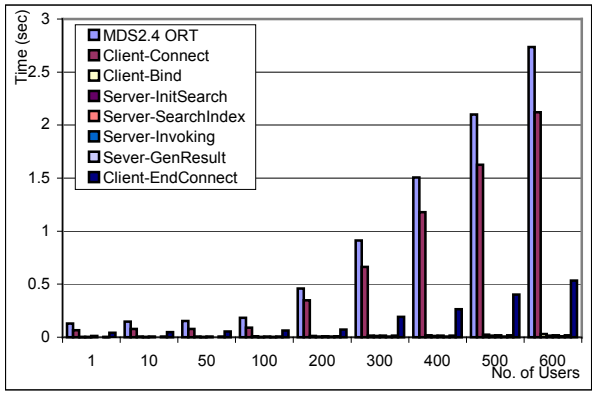
(1) MDS v2.4 GRIS (data in cache)

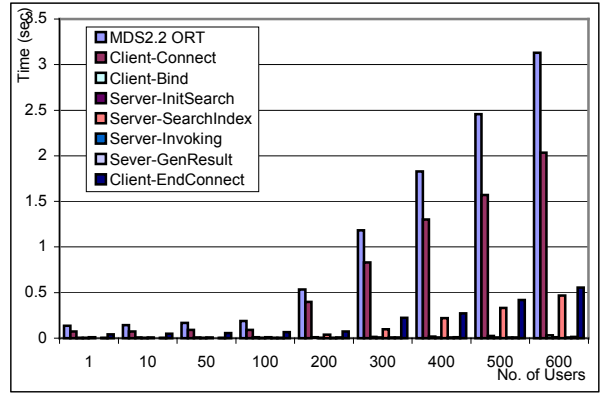
(2) MDS v2.2 GRIS (data in cache)

**Figure 6: MDS2 GRIS (data in cache) Phases Performance vs. No. of Concurrent Users**

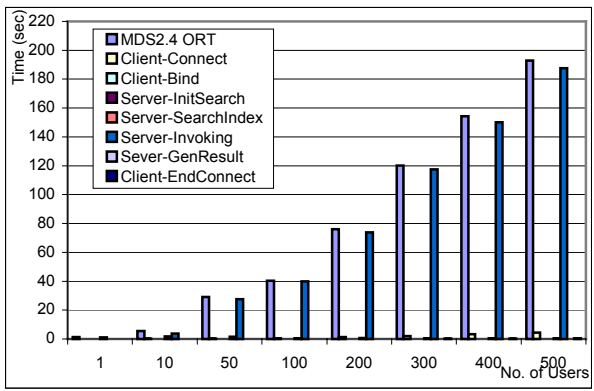
(1) MDS v2.4 GRIS (data not in cache)

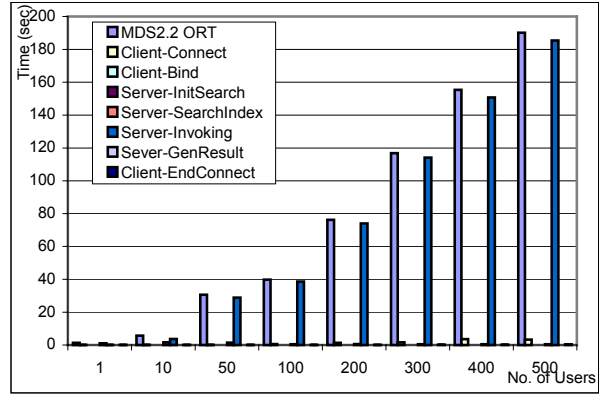
(1) MDS v2.2 GRIS (data not in cache)

**Figure 7: MDS2 GRIS (data not in cache) Phases Performance vs. No. of Concurrent Users**

Figures 2–7 show the performance results of v2.2 and v2.4 MDS2 GRISes, for two scenarios described above. The end-to-end performance results are presented in Figures 2–5 (Figures 2 compares v2.4 GRIS throughput performance with v2.2 GRIS while the rest figures present v2.4 results only), while Figures 6 and 7 show the NetLogger instrumentation results about the performance of the seven phases constituting the end-to-end path.

The results show that a GRIS, if configured with data in cache, can achieve a much higher scalability and end-to-end performance (throughput performance in Figure 2 and *ORT* performance Figure 3, respectively) than one without data caching, as one would expect. Comparing the performance results of the divided phases of end-to-end path for each scenario (shown in Figures 6 and 7), we found that the *RPT* occupies more than 90% of the *ORT* when a GRIS doesn't cache data. The much longer delay in the *Server-Invoking* phase is the source of the degraded performance. Since *Server-Invoking* is the stage in which a GRIS invokes the reported information providers to get the data, we believe the delay is caused by the fact that the cost to execute information providers can be expensive. To make the delay even worse, concurrent queries asking information from the same information provider must compete with each other, since a GRIS can only serve them serially.

For the GRIS with data in cache, the *ORT* did not exceed 3 seconds, compared with a maximum *ORT* for a GRIS without data caching of 190 seconds. The reason is a GRIS can serve the concurrent queries with data in its cache rather than invoking low-level information providers, and all cached data can reside in memory to further improve the efficiency. Figure 6 confirms the *Server-Invoking* phase, and the *RPT* is no longer the source of performance bottleneck. However, we observe that the throughput did not follow a constantly increasing rate after the point of 200 concurrent users for both versions of GRIS (Figure 2).

This effect is the result of the longer delay of the *Client-Connect* time (Figure 6). Unlike the GRIS without data in cache, the performance depends on the *Client-Connect* time – the sum of *Client-Connect* time and *Client-EndConnect* time is about 95% of *ORT*. Many factors may attribute to the delay of *Client-Connect*, for example, network limitations at the server side, the operating system's scheduling delay, or the constraint of the LDAP protocol used by MDS2.

Generally, a GRIS experienced a higher load (load1 results in Figure 4 and CPU_load results in Figure 5) with the increasing number of users no matter it caches data or not for each version. This is because more concurrent queries contest for CPU to acquire the service of the GRIS. However, the machine hosting a GRIS without data in cache presents a lower load than hosting a GRIS that caches data, indicating that many of the processes were blocked waiting for resources in the former case.

From the performance results, we also observed differences between v2.4 GRIS and v2.2 GRIS. Figure 7 shows v2.4 GRIS outperforms v2.2 GRIS in the efficiency of processing requests. The reason for the improved performance is that v2.4 GRIS spends less time on the *Server-SearchIndex* phase than does v2.2 GRIS, especially with a large number of users. This is likely due to better memory use in v2.4. We conclude that the overhead for the MDS2 GRIS can be substantially reduced by data caching since invoking the information providers to serve each query can be expensive We suggest that, in order to provide good quality of service, a GRIS should always cache the data that is static or is expensive to calculate or fetch. Moreover, a GRIS should support fewer than a 100 users if it has to provide fresh data without data caching for each query.

### 3.4. MDS2 Directory Server Scalability

The second functionality of MDS2 we tested was the performance of GIIS as a directory server with the number of concurrent users.

We ran each version of MDS2 GIIS on *lucky1* with a GRIS (containing information from 10 information providers) on each of *lucky3–7* registered to it. To analyze only the directory functionality of the GIIS and not its information serving capacity as an aggregation server, we set the *cachettl* (cache element time to live) parameter to a value larger than 600 seconds to make sure the data was always in the cache during each round of the experiments. Each user queries for all the data elements from the GIIS directory. This means the average data size a query expects is approximately five times bigger than that in the GRIS experiments, about 50 KB.

Figures 8–12 show the performance results of v2.2 and v2.4 MDS2 GIISes with data in their caches. The end-to-end performance results are presented in Figures 8–11, while Figure 12 shows the phase performance instrumented by NetLogger.

With data in cache, MDS2 GIIS scales well and exhibits a high throughput and low *ORT* with respect to the increasing number of users. These results are due to the fact MDS2 GIIS is very efficient in processing the queries at the server side (the *RPT* was always smaller than 0.2 sec, shown in Figure 12) since it does not need to communicate with all the registered, lower-level GRIS to generate the fresh data. We can expect the communication expense is nontrivial, since GIIS and the registered GRIS run on different machines.

The NetLogger instrumentation results shown in Figure 12 also illustrate that the majority of *ORT* is spent on the client side's *Client-Connect* phase for MDS2 GIIS. More concurrent users accessing the same GIIS simply means each user will experience a longer latency in building the connection to the GIIS service on average. Since MDS2 GIIS and GRIS are constructed on nearly the same underlying protocols, we attribute the longer delay of *Client-Connect* time to the same reason we gave to GRIS.

The performance difference of different versions of MDS2 GIIS is also reflected in the results. The v2.4 GIIS shows a higher throughput (Figure 8) and lower *ORT* (Figure 9) than does the v2.2 GIIS when they are accessed by a same number of users. The probable explanation is better use of memory.

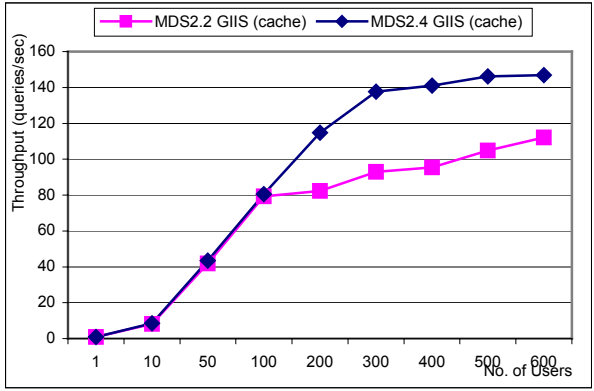

**Figure 8: GIIS (data in cache) Throughput vs. No. of Concurrent Users**

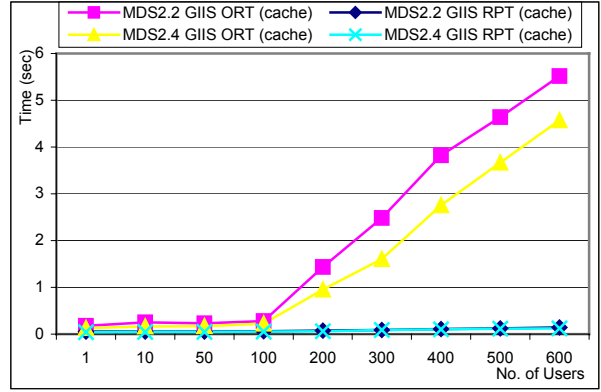

**Figure 9: GIIS (data in cache) *ORT* and *RPT* vs. No. of Concurrent Users**

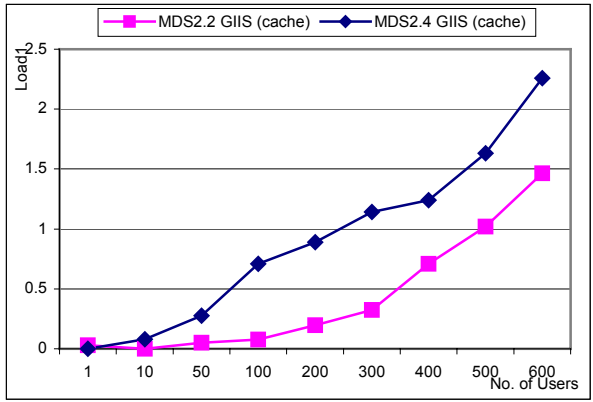

**Figure 10: GIIS (data in cache) Host Load1 vs. No. of Concurrent Users**

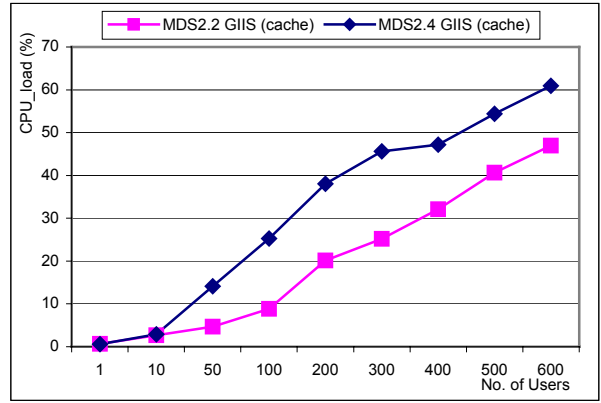

**Figure 11: GIIS (data in cache) CPU_load vs. No. of Concurrent Users**

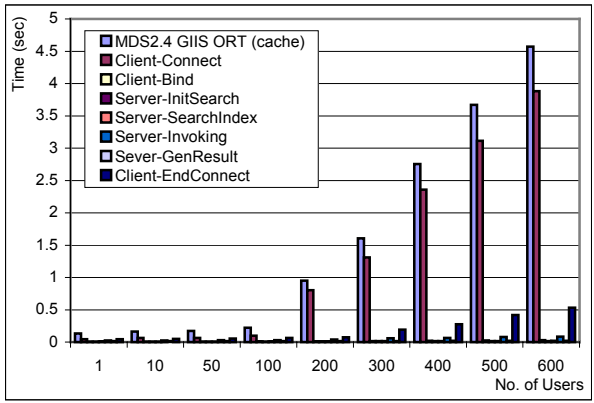

(1) MDS v2.4 GIIS (data in cache)

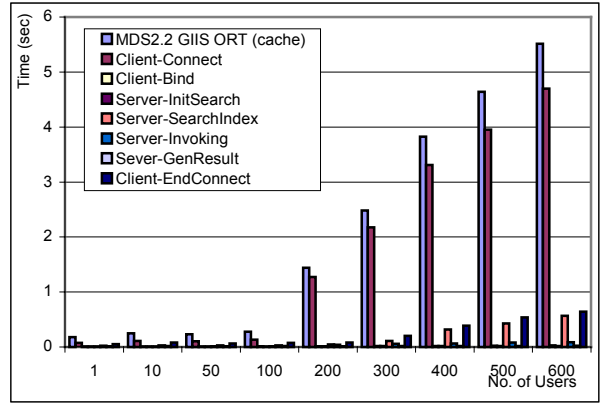

(2) MDS v2.2 GIIS (data in cache)

**Figure 12: MDS2 GIIS (data in cache) Phases Performance vs. No. of Concurrent Users**

Though MDS2 GIIS with data in cache can be treated similarly to a GRIS with data in cache, their absolute performance is quite different. When accessed by the same number of users, a GRIS is more efficient in serving queries than is the same version of GIIS because the GIIS has many more entries and the searching takes longer.

From the above experiment we see that using the MDS2 GIIS as a directory server with data caching is a good choice. It can provide good quality of service if serving fewer than 400 users concurrently. With a larger number of users, however, one should duplicate the GIIS in order to keep the quality of service.

## 4. Conclusion

In this paper, we have investigated the scalability and performance of the Globus Toolkit MDS2. We used NetLogger to instrument the MDS2 codes at both the server side and client side, and we broke the end-to-end path of a user query into seven phases to diagnosis the performance bottlenecks. Our work shows that, when accessed by a large number of concurrent users, both MDS2 GRIS and GIIS present good scalability and performance if they keep data in cache. On the other hand, their performance degrades dramatically without data caching. The NetLogger instrumentation results show that a primary cause of the poor performance is either invoking the reported information provider or consulting the reported GRIS. We also find that the primary components of Grid middleware must be available at well-connected sites, because of the high load seen in the experiments we evaluated.

In our future work, we plan to do more experiments to address other characteristics of MDS2 GRIS and GIIS with NetLogger instrumentation, for example, how the performance of a GRIS scales with the amount of data it contains. We also plan to compare the MDS2 performance with other Grid middleware in the same category, such as R-GMA[1] and Hawkeye[7].

## Acknowledgments

We thank both John Mcgee and Ben Clifford at ISI, for assistance with the MDS2; and both Brian Tierney and Dan Gunter at LBNL, for assistance with NetLogger. We also thank Scott Gose and Charles Bacon for assistance with the testbed at Argonne. This work was supported in part by the Mathematical, Information, and Computational Sciences Division subprogram of the Office of Advanced Scientific Computing Research, U.S. Department of Energy, under contract W-31-109-Eng-38.